\documentclass[structabstract]{aa}%
\usepackage[english]{babel}%
\usepackage[dvips]{graphicx}%
\usepackage{amssymb}
\usepackage{hhline}

\begin{document}

\title{Hubble flow around Fornax cluster of galaxies}

\author{O.\,G. Nasonova\inst{1}\fnmsep\inst{2}\fnmsep\thanks{N\'ee Kashibadze.}\and
        J.\,A. de Freitas Pacheco\inst{1} \and
        I.\,D. Karachentsev\inst{2}\fnmsep\inst{3}}

\institute{Universit\'e de Nice -- Sophia Antipolis, Observatoire de la C\^ote d'Azur\\
          Laboratoire Cassiop\'ee, UMR 6202, BP-4229, 06304,  Nice Cedex 4, France\\
\and
          Special Astrophysical Observatory of the Russian Academy of Sciences\\
          Laboratory of Extragalactic Astrophysics and Cosmology\\
          Nizhnij Arkhyz, Karachay-Cherkessia, 369167, Russia\\
\and
          Universit\'e de Lyon, Universit\'e Lyon 1, CNRS/IN2P3,\\
          Institut de Physique Nucl\'eaire de Lyon, Villeurbanne, France}

\date{\today}

\abstract
{}
{This work aims to provide a new mass estimate for the Fornax cluster and the
Fornax-Eridanus complex, avoiding methods like the virial or fits of X-ray
emission profile, which assume that the system is in equilibrium, probably not
the case of Fornax, still in process of formation.}
{Our mass estimate is based on the determination of the zero-velocity surface
which, in the context of the spherical infall model permits an evaluation of the
total mass inside such a surface. The zero-velocity surface radius $R_0$ was
estimated either by a running median procedure or by fitting the data to the
velocity field expected from the spherical model, including effects of the
cosmological constant. The velocity field in a region within 20 Mpc of the
Fornax center was mapped using a list of 109 galaxies whose distances have an
average accuracy of $0.31$ mag in their distance modulus.}
{Our analysis indicates that the mass of the Fornax cluster itself inside a
radius of $[2.62-5.18]$~Mpc is $[0.40-3.32]\times 10^{14}~M_\odot$ while the
mass inside $[3.88-5.60]$~Mpc, corresponding to the Fornax-Eridanus complex is
$[1.30-3.93]\times 10^{14}~M_\odot$.}
{}

\keywords{galaxies -- clusters -- individual: Fornax}

\maketitle

\section{Introduction}

In the hierarchical scenario of structure formation, clusters of galaxies are
one of the largest structures observed in nature. Clusters have been assembled
relatively late in the history of the universe, being located in the
intersection of filaments constituting the cosmic web (Voit 2005). Clusters are
bound essentially by the gravitational action of the so-called dark matter with
the luminous (baryonic) component given only a minor contribution to the
gravitational potential of the system. Moreover, most of the baryonic matter is
under the form of hot and warm gas filling the intracluster medium, which is
detected by its X-ray emission. Clusters of galaxies have a particular interest
in cosmology since the evolution of their number density above a given mass
provide strong constraints on parameters characterizing different world models
(Rosati et al. 2002). 

Masses of clusters are generally estimated by using the virial relation, which
presupposes that the system is dynamically relaxed. This is not the case during
phases in which the cluster undergoes a merger episode or accretes mass through
filaments (see, for instance, Girardi \& Biviano 2002). Moreover, this method is
affected by the eventual inclusion of interlopers by projection effects in the
considered sample and by dynamical friction, which may introduce an important
bias between the velocity dispersion of galaxies and dark matter particles. The
reliability of the virial relation as a mass estimator has been checked by
numerical simulations performed by different authors. Danese et al. (1981) and
Fernley \& Bhavsar (1984) concluded that projection effects are important and
may affect considerably the virial masses. Similar results were obtained by
Perea et al. (1990), who reached the conclusion that the virial, in general,
overestimates the masses unless interlopers are eliminated. A more recent study
by Biviano et al. (2006) led to more quantitative results: the virial relation
overestimates true masses by about 10\% if the simulated clusters have more than
60 members, with uncertainties increasing up to 50--60\% for objects having
15--20 members.

Besides the virial, other methods have been frequently employed as mass
estimators of clusters. The X-ray emission profile of the hot intracluster gas
can be used to trace the gravitational potential of clusters, under the
assumption of hydrostatic equilibrium (Sarazin 1986; Reiprich \& Bohringer
2002). The accuracy of this approach was tested by cosmological simulations,
which indicate that when masses are evaluated inside a radius at which the mean
cluster density is about 500--2500 times the critical density, the uncertainties
are of the order of 14--29\% (Evrard et al. 2006). The masses of galaxy clusters
can be also estimated through the analysis of gravitational lensing, since the
gravitational field of clusters distorts the image of galaxies situated behind
them (Broadhurst et al. 1995). A comparison of masses resulting from X-ray data
and strong lensing shows that values drawn from the latter method are, on the
average, twice those obtained from the former procedure (Wu 2000). These
differences could be the consequence of an oversimplification of the lens model
and/or the violation of the hypothesis of the gas isothermality. More recently,
weak gravitational lensing, a technique permitting to track the gravitational
potential of these objects by the distortion induced in the shape of background
galaxies (see, for instance, Mellier 1999) has been used to estimate masses of
different clusters. However, the presence of nearby filaments leads in general
to an overestimate of masses by 10--30\% (Metzler et al. 1999).

All the above mentioned methods correspond to scales less or of the order of
1--2 Mpc, which are typical dimensions of the central region of galaxy clusters.
At larger scales, corresponding to the surroundings of clusters, galaxies are
probably falling onto the cluster for the first time. These outskirt galaxies
despite of being bound to the cluster are not in dynamical equilibrium. In this
case, the knowledge of the velocity field of these objects may lead to an
estimate of the central region mass. In fact, such an approach was proposed by
Lynden-Bell (1981) and Sandage (1986) based on the spherical infall model. The
motion of the outskirt galaxies is supposed to be essentially radial and from
the knowledge of the distance $R_0$, at which the radial velocity with respect
to the center of mass is zero, the mass inside such a surface can be estimated.
The spherical model predicts also the existence of caustics, surfaces at which
(theoretically) the galaxy number density is infinite (Regos \& Geller 1989).
The profile of the caustic amplitude, as seen in a phase-space diagram for the
outskirt galaxies, can be used as a mass estimator with an accuracy of a factor
of two (Diaferio 1999).

In the present paper we intend to present a new estimate of the mass of the
southern cluster located in Fornax (Abell~S0373) at a distance of 20 Mpc. An
early survey in the Fornax area performed by Ferguson (1989) indicates that
probably 340 galaxies are cluster members and a fit of the projected density
with a King profile suggests a core radius of about $0.7^\circ$. Despite of
being a cluster less rich than Virgo, the system presents different interesting
features. Its main structure is centered in NGC\,1399. According to Drinkwater
et al. (2001), dwarf galaxies form a distinct population which is probably
infalling onto the main system. Using the method by Diaferio (1999) mentioned
above, which does't assume dynamical equilibrium, Drinkwater et al. (2001)
estimated the projected mass inside a radius of 1.4~Mpc as $(7 \pm 2)\times
10^{13}~M_{\odot}$. Since then a large amount of data on galaxies within the
Local Universe, including radial velocities and precise distances have been
accumulated. This justifies a novel study of the velocity field in the Fornax
region as well as new estimates of its mass, based on the analysis of the
velocity field of outskirt galaxies, modeled by the spherical infall model and
avoiding problems present in other mass indicators as mentioned above. This
paper is organized as follows: in Section 2 the available data is presented, in
Section 3 mass estimates are discussed and finally in Section 4 the main results
are summarized.

\section{The data}

In the past decade a significant number of galaxies present in the Local Volume
had their distances measured with a quite good accuracy, in particular thanks to
data obtained with the Hubble Space Telescope (Karachentsev et al. 2002a, 2002b,
2006, 2009). Besides galaxies present in the neighborhood of the Local Group a
substantial effort was also made to increase the database on the Virgo cluster.

Distances to galaxies in the Local Universe have been estimated from different
methods:

1) TRGB, based on the luminosity of the Tip of the Red Giant Branch, considered
as one of the most efficient methods to determine distances of nearby galaxies,
practically independent on their morphological type. The method requires images
in two ore more photometric bands obtained with WFPC2 or ACS cameras on board of
the HST, yielding an accuracy of about 7\% on distances derived by such a
procedure (Rizzi et al. 2007). A consolidated list of distances for galaxies in
the Local Volume is given in the {\itshape{}Catalog of Neighboring Galaxies}
(hereafter CNG, Karachentsev et al. 2004). Galaxies from CNG with only TRGB or
Cepheid distances were used, including some new determinations (Karachentsev et
al. 2006, Tully et al. 2006).

2) The surface brightness fluctuation method (SBF), applied to early type
galaxies, assumes that the old stellar population present in those objects gives
the main contribution to their luminosity. The method presupposes that the
brightness distribution is not affected by irregularities as, for instance, that
introduced by the presence of dust clouds. Using this approach, Tonry et al.
(2001) determined SBF distances for 300 E and S0 galaxies with typical errors of
$\sim$ 12\%. Galaxies in this sample are distributed over the whole sky,
extending up to cz $\sim 4000$~km~s$^{-1}$ and having a median velocity of
$\sim$1800~km~s$^{-1}$. 

3) Blakeslee et al. (2009) undertook a two-color ACS/HST imaging survey
including 43 early type galaxies situated in the Fornax core (the ACS Fornax
Cluster Survey project, hereafter ACS-FCS), deriving SBF distances with errors
of about 8\%. To the ACS-FCS list were added 18 dwarf ellipticals belonging to
the cluster, having SBF distances with an accuracy of about 9\% estimated by
Jerjen (2003) and Dunn \& Jerjen (2006). These authors suggest from the S-shaped
pattern distribution of these galaxies that Fornax is still in process of
formation.

4) Two galaxies within 15 Mpc of the Fornax cluster with distances measured with
an accuracy of 5\% by using SNIa light curves (Tonry et al. 2003) were also
included in our database.

5) Kashibadze (2008) determined distances for 402 edge-on spiral galaxies
selected from the {\itshape{}2MASS Flat Galaxy Catalog} (2MFGC, Mitronova et al.
2004), having radial velocities less than 3000~km~s$^{-1}$. Using a
multiparametric NIR Tully-Fisher relation, distances with an accuracy of about
$20$\% were obtained. The zero point of the {\itshape{}luminosity-line width}
relation was established by using 15 galaxies with distances derived from
cepheids and TRGB data.

\begin{figure*}
\centering
\includegraphics[height=17cm,keepaspectratio,angle=270]%
{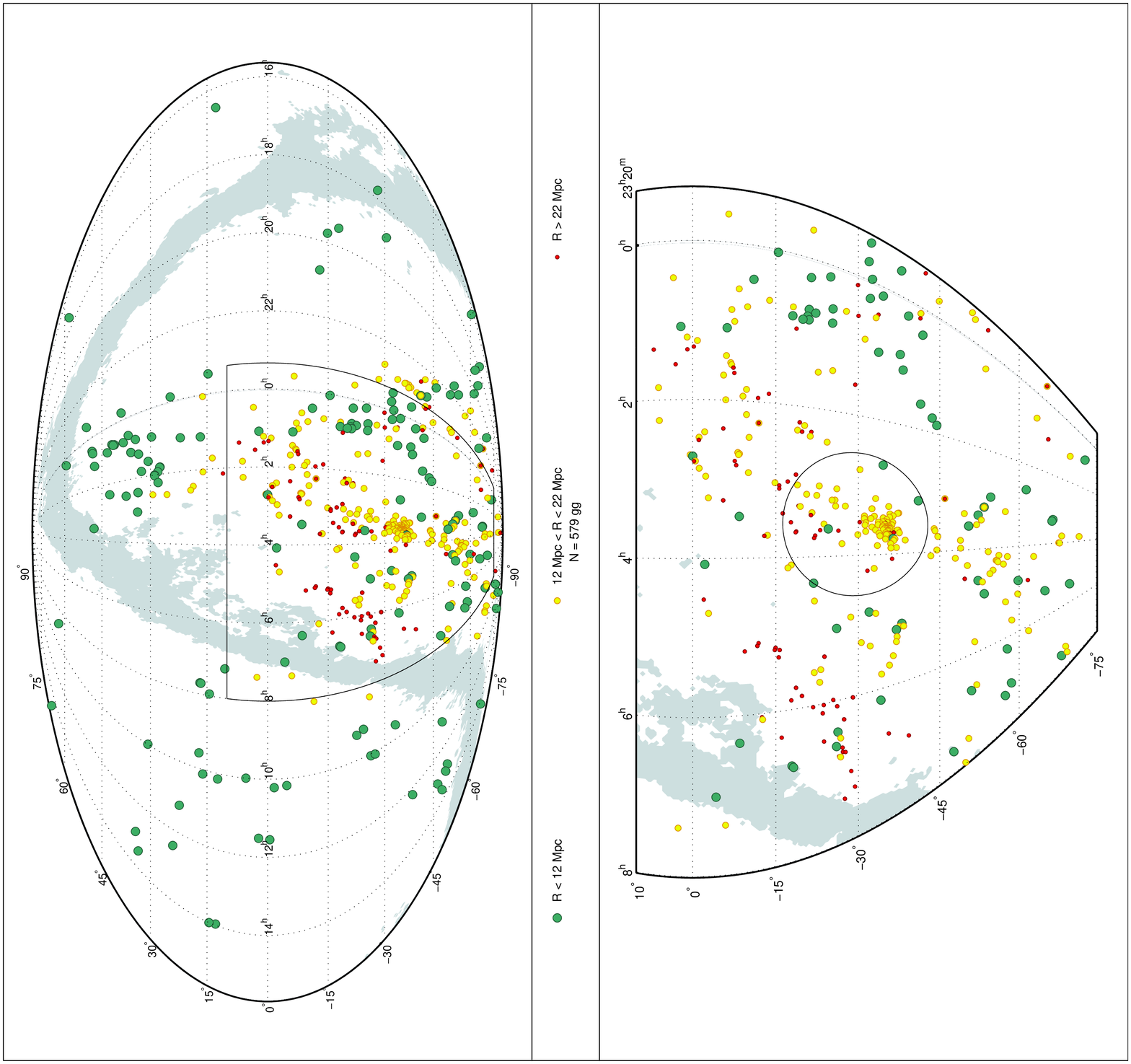}
\caption{Sky distribution in equatorial coordinates of 562 galaxies within 20
Mpc from the Fornax center. Markers of different diameters indicate three
distinct distance ranges from the Local Group. The circle in the lower panel
corresponds to the zero velocity sphere projection ($R_0 = 4.60$~Mpc) for the
Fornax-Eridanus complex.}
\end{figure*}

6) We supplemented the aforementioned samples with a compilation of distances by
Tully et al. (2008, 2009) derived from the Tully-Fisher relation calibrated for
optical ($B, V, R, I$) magnitudes. This compilation relays on numerous HI line
and photometric observations carried out by Methewson \& Ford (1996), Haynes et
al. (1999), Tully \& Pierce (2000), Koribalski et al. (2004), Springob et al.
(2005), Theureau et al. (2006) and other authors. Again, the zero point of these
relations were set by using a sample of 40 galaxies with distances determined by
cepheids and TRGB data. Finally, the list of galaxies by Springob et al. (2007)
(SFI++ sample), not considered by Tully et al. (2009) was also used in our
compilation.

The first four samples are referred as {\itshape{}precise data} because their
typical errors don't exceed 10--12\,\%, while the last two datasets are
mentioned as {\itshape{}Tully-Fisher data}.

In this paper we will generally follow Karachentsev \& Nasonova
(2010) and examine the {\itshape{}3D sample} considering galaxies with limited
spatial distances from the Fornax cluster center. As it was discussed by the
authors, this approach isn't free of systematical effects because galaxy
distances are measured with errors and their significance is different at the
proximate and the distant boundary of the spherical volume (the so-called
Malmquist bias; see Figure~2).

Our initial list includes 1140 galaxies within 30 Mpc of the Fornax cluster
(available in the electronic version of the paper). However, in the present
study, we will be focused in a region of 20 Mpc around the center of the
cluster, representing a sample of 562 objects which, in principle, would permit
an estimate not only of zero velocity surface but also of the transition region
between bound and unbound objects, the latter tracking essentially the Hubble
flow. The characteristics of these galaxies are given in Table~1. The first
column indicates the distance estimate method; the second column gives the mean
error in the distance modulus; the third column gives the number of galaxies and
the last column gives the sample goodness defined as
$G=(N/100)^{1/2}\cdot\sigma^{-1}_m$ (Kudrya et al. 2003). Figure~1 shows the
projected distribution of these galaxies in the sky. Galaxies are marked as
circles and their diameters indicate different distance ranges. 

\begin{table}[]
\caption{Database for galaxies within 20 Mpc from the Fornax center.}
\begin{center}
\begin{tabular}{lcrcrcr}
\hline
Distance method & $\sigma_m$ &  N  &   G  \\
\hline
TRGB + Ceph & 0.15 & 113 &  7.1 \\
SBF (Tonry) & 0.25 &  69 &  3.3 \\
ACSFCS+     & 0.19 &  55 &  3.9 \\
SNIa (Tonry)& 0.10 &   2 &  1.4 \\
TF (IR)     & 0.40 &  69 &  2.1 \\
TF (opt)    & 0.40 & 254 &  4.0 \\
\hline
All         & $-$  & 562 & 27.8 \\
\hline
\end{tabular}
\end{center}
\end{table}

\subsection{The velocity field}

In order to map the velocity field around the Fornax center and to have an
estimate of the {\itshape{}velocity-distance} relation, accurate radial
velocities and distances are required. Then, in the next step, these data must
be converted into distances and velocities relative to the cluster center.

Data on radial velocities, mostly from HI observations, were obtained in most
cases from the same sources of distances. When not available, NED data on
heliocentric velocities were used. Observational errors in radial velocities are
quite small (1--2~km~s$^{-1}$) in the case of HI observations (Tifft \&
Huchtmeier 1990) and can be neglected in comparison with distance errors
($\Delta V \ll \Delta R \cdot H_0$) in the scales of the nearest clusters ($R
\gtrsim 15$\,Mpc).

The transformation of heliocentric velocities into the Local Group reference
frame was performed with the standard apex parameters (Karachentsev \& Makarov
1996) adopted in NED. If $\varphi$ is the angular separation between the apex
and a galaxy, then the converted velocity is $V_{con} = V + V_{apex}\cos
\varphi$ and the error of this conversion is not more than $\left[(\Delta V)^2 +
(\Delta V_{apex})^2 + (\Delta \varphi \cdot V_{apex})^2\right]^{1/2}$, where
$\Delta V_{apex} = 5$~km~s$^{-1}$ and $\varphi \approx 1$\%, so $\Delta \varphi
\cdot V_{apex} \approx 3$~km~s$^{-1}$. Thus, the errors introduced by this
transformation are about 6~km~s$^{-1}$, being negligible in comparison with
distance errors.

The gravitational effect of the Fornax cluster can be seen directly from the
radial velocity vs. distance relation as an S-shaped wave. Radial velocities and
distances relative to the Local Group centroid for 98 galaxies in the cluster
core ($\theta < 5^\circ$) are represented in the top panel of Figure~2. Here,
precise distances for most galaxies were obtained within the special survey
ACS-FCS with HST (Blakeslee et al. 2009). The centroid of galaxies forming the
``virial column'' at [$19.58\pm1.25$] Mpc is marked by gray (see Section~3.1 for
discussion of the Fornax cluster barycenter position). The plotted value of
virial radius, $\pm1.25$~Mpc, is an approximate estimate based on the $R_0$
value for the Fornax-Eridanus complex obtained in this paper (4.60~Mpc). The
S-shaped curves correspond to a Hubble flow perturbed by a point-like masses of
$1.30\cdot10^{14}M_{\odot}$ and $3.93\cdot10^{14}M_{\odot}$ as the limiting
cases within the confidence range (see Section 3.2 for details) in the case of
line-of-sight passing exactly through the cluster center. The typical distance
error bars for datasets (2), (3) and (4) are shown.

The distribution of radial velocities and distances for remaining galaxies of
the sample in periphery of the Fornax cluster ($5^\circ < \theta<30^\circ$) is
shown in the bottom panel of Figure~2. Here, the S-shaped lines having lower
amplitudes describe the behavior of perturbed Hubble flow at angular distance
$\theta=5^\circ$. The typical distance error bars for datasets (1), (6) and (7)
are presented.

\begin{figure*}
\centering
\includegraphics[height=13cm,keepaspectratio,angle=270]{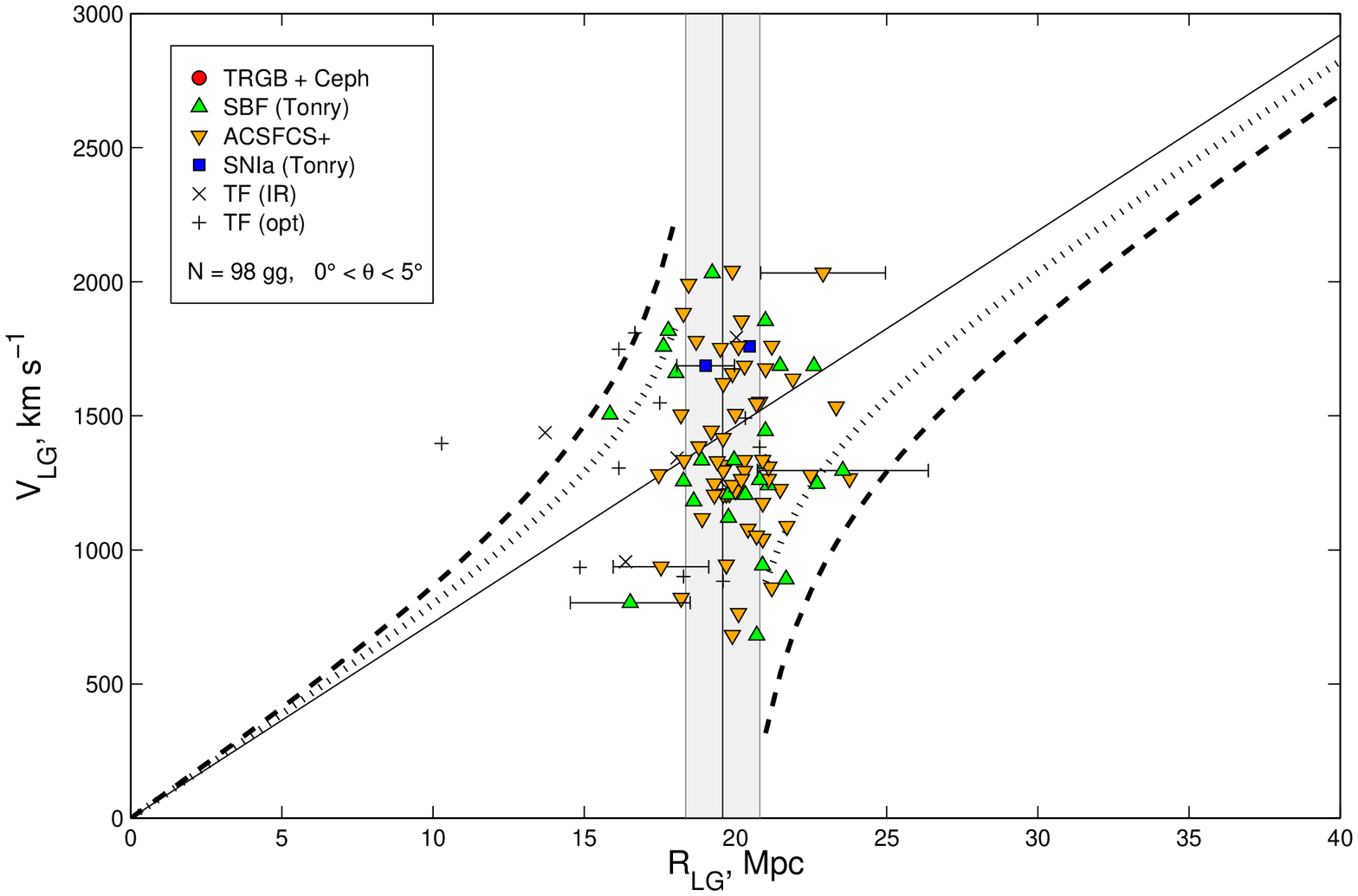}
\includegraphics[height=13cm,keepaspectratio,angle=270]{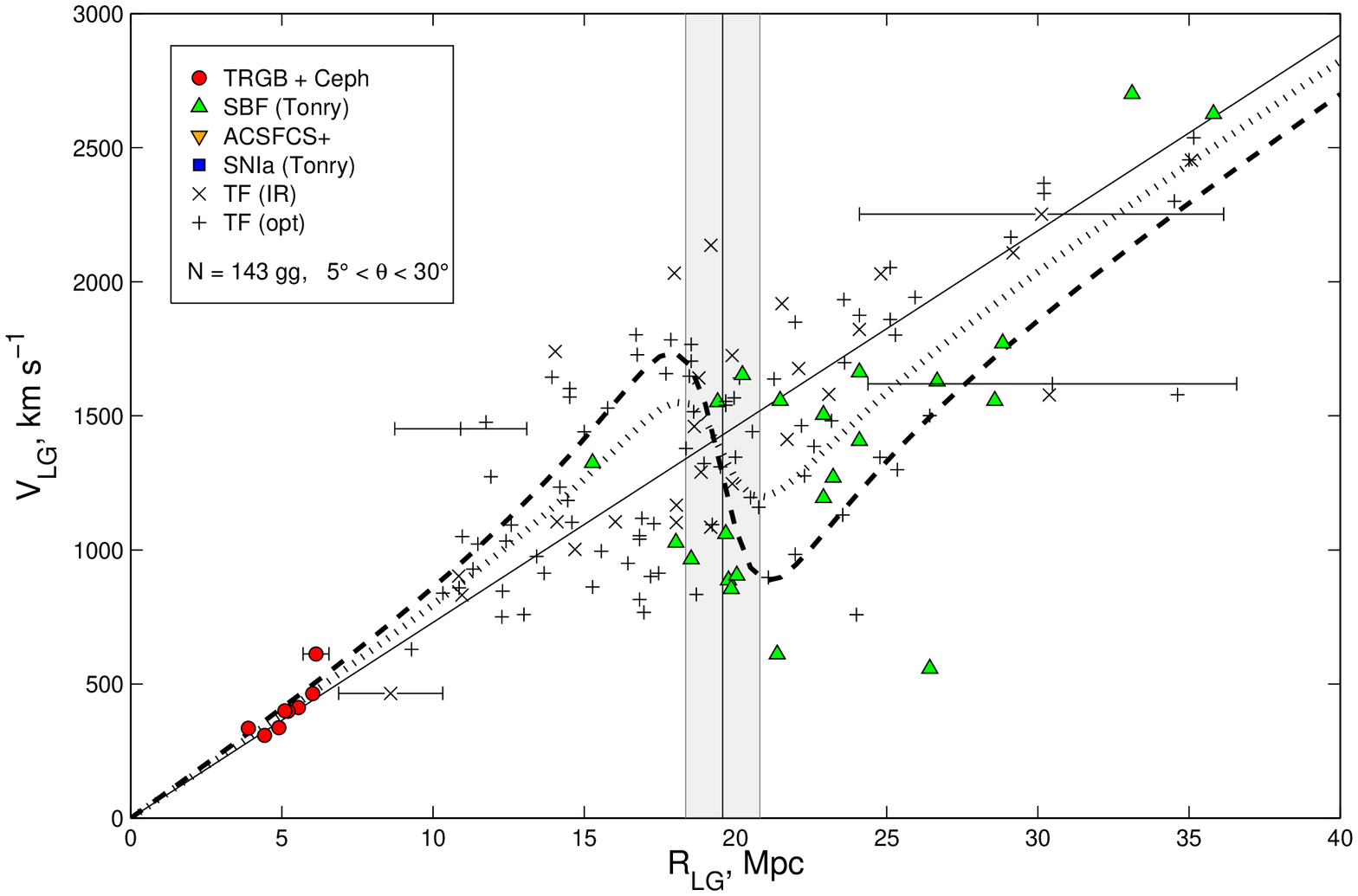}
\caption{Radial velocity vs. distance relation for galaxies in the Fornax
cluster region with respect to the Local Group centroid. Galaxy samples with
distances derived by different methods are marked by different symbols. The
inclined straight line traces Hubble relation with the global Hubble parameter
$H_0=73$~km~s$^{-1}$~Mpc$^{-1}$. The virial zone is filled with gray. Two
S-shaped lines correspond to a Hubble flow perturbed by masses of $1.30 \cdot
10^{14}M_{\odot}$ (dotted) and $3.93 \cdot 10^{14}M_{\odot}$ (dashed) as the
limiting cases within the confidence range. The typical distance error bars for
each dataset are shown. {\itshape{}Top}: the cluster core within angular
distance $\theta < 5^\circ$, the S-shaped lines indicate the expected infall at
$\theta = 0^\circ$. {\itshape{}Bottom}: peripheric regions with $5^\circ <
\theta < 30^\circ$, the S-shaped lines indicats the infall at $\theta =
5^\circ$.}
\end{figure*}

However, a serious source of uncertainty is caused by the absence of data on
tangential velocities. In order to estimate galaxy velocities with respect to
the center of mass of Fornax, some model should be used and thus, the results
turn out to be model-dependent. Taking into account the lack of data on the true
velocity vector of galaxies, there are at least two approaches to obtain such a
transformation.

The first one assumes that the prevailing motion, which involves most of the
galaxies under study is the asymptotic Hubble relationship (the model of the
{\itshape{}minor attractor}). The second approach considers that galaxies are
within the infall zone (the model of the {\itshape{}major extended attractor}),
i.e., they do not follow the Hubble flow but instead are falling towards the
cluster center. Both cases were discussed in details by Karachentsev \& Nasonova
(2010); see Figure~3 sketching the relative positions of the considered galaxy
($G$), the observer ($LG$) and the cluster center).

\begin{figure*}
\sidecaption{}
\includegraphics[height=12cm,keepaspectratio,angle=270]{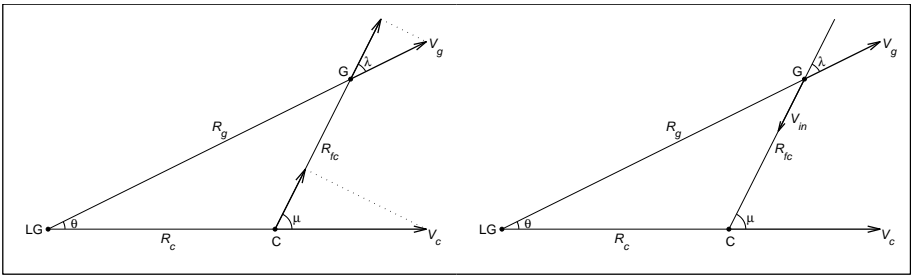}
\caption{Relative positions between the observer $LG$, the considered galaxy $G$
and the cluster center $C$. {\itshape Left panel}: Case of a pure Hubble flow.
{\itshape Right panel }: Case of a pure infall towards the Fornax center.
\vspace{-3.7cm}}
\end{figure*}

When a galaxy is located strictly in front or behind the cluster center (i.e.
the angles $\lambda$ and $\theta$ are small), both approaches yield the same
infall velocity toward the cluster center. When $\lambda$ is close to
90$^\circ$, in the second case $V_{in} \to \infty$ leading to a significant
discrepancy between the two approaches. However, there are no dramatic
differences between both methods in the Hubble diagram. Some galaxies move along
the vertical axis appreciably, but the behavior of running medians (see next
section) traces the infall of galaxies towards the cluster in a similar way.
Nevertheless, as we shall see later, the second method yields systematically
slightly larger values of $R_0$. The scatter of galaxies in the Hubble diagram
also increases in the second case. These considerations suggest the first
approach to be preferred.

\begin{figure*}
\sidecaption
\includegraphics[height=12cm,keepaspectratio,angle=270]{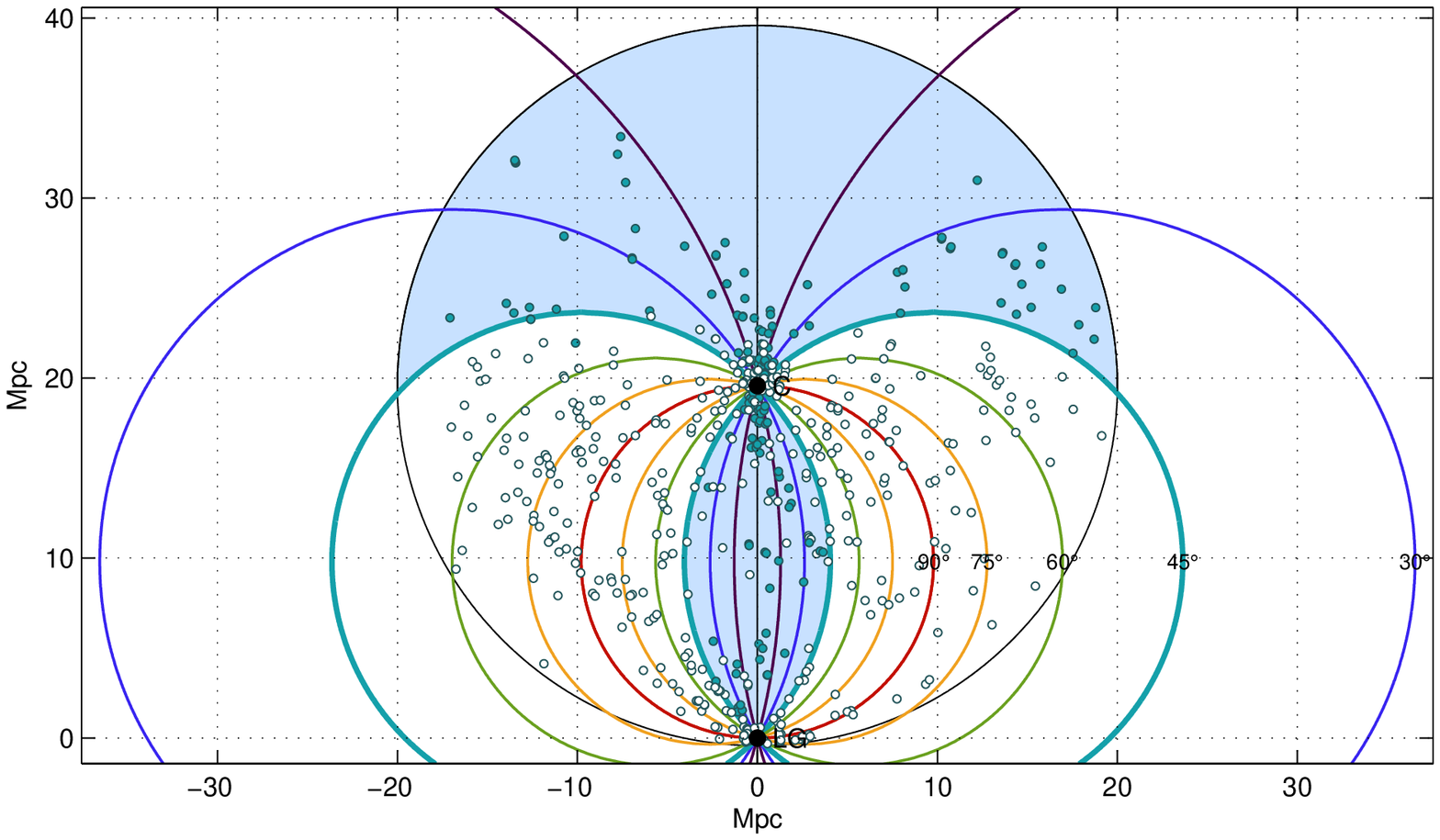}
\caption{Spatial distribution of galaxies around the Fornax cluster. The zones
where $\lambda < 45^\circ$ or $\lambda > 135^\circ$ are blue shaded whereas the
exclusion zones are white. The same color code indicates galaxies included or
not in the analysis. For a question of clearness, lines of equal $\lambda$ for
values in the range $15^\circ$--\,\,$90^\circ$ are also shown.
\vspace{-6.3cm}}
\end{figure*}

Finally, in order to reduce the role of the unknown tangential component of the
velocity and to avoid further uncertainties, we decided to select for our
analysis only galaxies situated approximately in front and behind the cluster,
i.e. in a cone satisfying the conditions $\lambda<45^\circ$ or
$\lambda>135^\circ$. The number of galaxies satisfying this additional
constraint within 20 Mpc of the Fornax center is 164 and their projected
distribution in the sky is shown in Figure~4.

It should be mentioned that the role of possible chaotic tangential velocities
of the galaxies had been studied by Karachentsev \& Kashibadze (2006). They have
performed numerical simulations, adding some random tangential component to the
observed radial velocity. Their modeling of the Hubble flow in the vicinity of
the Local Group showed that typical tangential velocities with amplitudes of 35
and 70~km~s$^{-1}$ produce a statistical uncertainty in the evaluation of the
zero-velocity surface radius as small as $\pm$2\% and $\pm$4\% respectively.

\section{Mass estimates}

The difficulties with the different mass estimators were already mentioned. In
particular, the presence of interlopers lead to an overestimate of the mass
ranging from 10\% up to 60\% when the virial relation is used, whereas estimates
based on X-ray data have uncertainties ranging from 14--30\%.

\begin{figure}
\resizebox{\hsize}{!}{\includegraphics[angle=270]{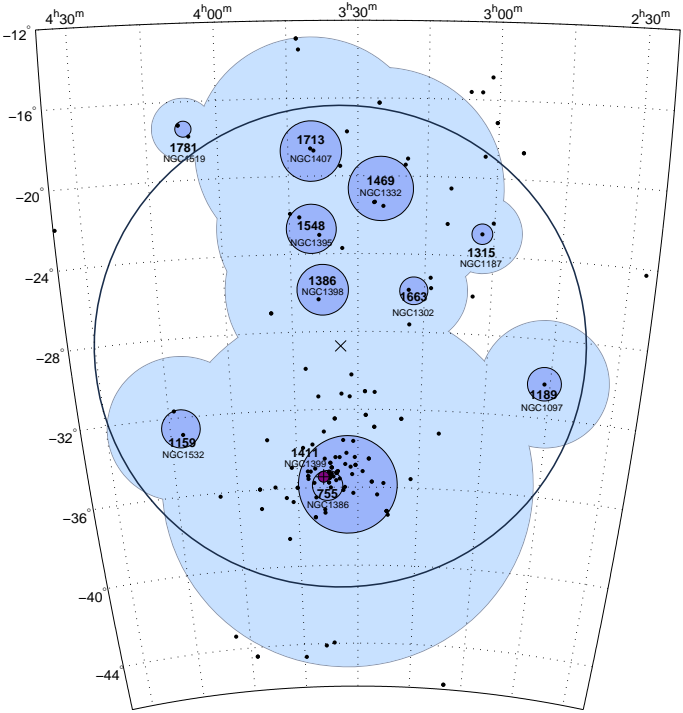}}
\caption{Fornax cluster ({\itshape{}NGC\,1399 group}), Eridanus subcluster
({\itshape{}NGC\,1395, NGC\,1407, NGC\,1332, NGC\,1398 groups}) and other
neighboring groups of galaxies forming the Fornax complex (Makarov \&
Karachentsev, 2011). In the scales of virial radii (dark-blue circles) these
overdensities constitute separate bound groups, but they overlap substantially
in scales of $R_0$ (light-blue regions). The velocities of groups
$V_\mathrm{LG}$ are indicated near the names of their central galaxies.
NGC\,1399 as the most massive galaxy in the complex situated near the center of
the hot X-ray emitting gas is indicated as a purple circle marked with a cross.
Black dots represent galaxies with measured distances. The large bold circle of
radius $13.2^\circ$ outlines the zero-velocity sphere ($R_0 = 4.60$~Mpc) around
the Fornax-Eridanus cluster barycenter.}
\end{figure}

The case of the Fornax cluster is quite particular. Unlike clusters as Coma or
Virgo, which have a well defined center and a more or less smooth mass
distribution, despite the presence of some substructures, the Fornax cluster has
a complicated mass distribution, including many substructures. Such a complexity
is sketched in Figure~5 where it is possible to identify: the Fornax cluster
centered on NGC\,1399, the Eridanus subcluster, including the groups NGC\,1407,
NGC\,1332, NGC\,1395 and NGC\,1398 as well as other small neighboring groups. It
should be emphasized that the virial radii of these groups do not overlap,
suggesting that these objects are separated and constitute gravitationally bound
structures. However these groups overlap in the scale of the zero-velocity
radius, indicating that despite the absence of dynamical equilibrium, they are
probably bound to the main structure.

\subsection{The spatial position of the Fornax-Eridanus barycenter}

\begin{table*}
\centering
\caption{Groups, triplets and pairs around the Fornax-Eridanus
complex with individual distances.}
{\tabcolsep=0.8ex
\begin{tabular}{|lcrrrrrrrrrrrrrrr|}
\hhline{|-----------------|}
\multicolumn{1}{|l}{Group}&
\multicolumn{1}{c}{RA, Dec}&
\multicolumn{1}{c}{$N$}&
\multicolumn{1}{c}{$N_D$}&
\multicolumn{1}{c}{$V_\mathrm{LG}$}&
\multicolumn{1}{c}{$\pm$}&
\multicolumn{1}{c}{$\sigma_v$}&
\multicolumn{1}{c}{$R_h$}&
\multicolumn{1}{c}{$\lg L_K$}&
\multicolumn{1}{c}{$\lg M$}&
\multicolumn{1}{c}{$D$}&
\multicolumn{1}{c}{$\pm$}&
\multicolumn{1}{c}{$D_h$}&
\multicolumn{1}{c}{$\theta$}&
\multicolumn{1}{c}{$\lambda$}&
\multicolumn{1}{c}{$\theta'$}&
\multicolumn{1}{c|}{$\lambda'$}\\
\multicolumn{1}{|l}{}&
\multicolumn{1}{c}{\itshape{}hh\,mm\,ss~dd\,mm\,ss}&
\multicolumn{1}{r}{}&
\multicolumn{1}{r}{}&
\multicolumn{2}{c}{km/s}&
\multicolumn{1}{r}{km/s}&
\multicolumn{1}{c}{kpc}&
\multicolumn{1}{c}{$L_\odot$}&
\multicolumn{1}{c}{$M_\odot$}&
\multicolumn{2}{c}{Mpc}&
\multicolumn{1}{c}{Mpc}&
\multicolumn{1}{c}{$^\circ$}&
\multicolumn{1}{c}{$^\circ$}&
\multicolumn{1}{c}{$^\circ$}&
\multicolumn{1}{c|}{$^\circ$}\\
\hhline{|-----------------|}
\multicolumn{1}{|l}{(1)}&
\multicolumn{1}{c}{(2)}&
\multicolumn{1}{r}{(3)}&
\multicolumn{1}{r}{(4)}&
\multicolumn{1}{c}{(5)}&
\multicolumn{1}{c}{(6)}&
\multicolumn{1}{r}{(7)}&
\multicolumn{1}{r}{(8)}&
\multicolumn{1}{c}{(9)}&
\multicolumn{1}{c}{(10)}&
\multicolumn{1}{c}{(11)}&
\multicolumn{1}{c}{(12)}&
\multicolumn{1}{c}{(13)}&
\multicolumn{1}{r}{(14)}&
\multicolumn{1}{r}{(15)}&
\multicolumn{1}{r}{(16)}&
\multicolumn{1}{r|}{(17)}\\
\hhline{|-----------------|}
NGC\,1097 & 024621.0$-$314430 &   4 &  1 & 1189 & 81 & 140 &  43 & 11.26 & 12.53 & 15.9 & 3.2 & 16.3 & 11 & 131 & 10 & 142\\
NGC\,1187 & 030238.6$-$230819 &   4 &  1 & 1315 & 28 &  48 & 106 & 10.60 & 11.90 & 18.4 & 3.7 & 18.0 & 15 &  96 &  9 & 125\\
NGC\,1201 & 030408.0$-$275549 &   2 &  2 & 1633 & 48 &  48 & 312 & 10.95 & 11.94 & 21.0 & 0.8 & 22.4 & 12 &  65 &  7 &  88\\
NGC\,1232 & 030945.5$-$212514 &   3 &  2 & 1582 & 49 &  70 & 475 & 11.07 & 12.93 & 19.2 & 0.5 & 21.7 & 16 &  86 & 10 & 113\\
NGC\,1291 & 031718.6$-$425331 &   3 &  3 &  703 & 70 &  99 & 226 & 11.03 & 13.09 &  8.2 & 0.5 &  9.6 &  6 & 170 & 13 & 159\\
NGC\,1302 & 031723.8$-$260702 &   6 &  4 & 1663 & 21 &  47 & 383 & 11.15 & 12.30 & 19.2 & 1.0 & 22.8 & 10 &  91 &  5 & 136\\
NGC\,1332 & 032500.7$-$212102 &  22 &  9 & 1469 & 40 & 183 & 279 & 11.55 & 13.39 & 24.1 & 0.8 & 20.1 & 15 &  45 &  8 &  43\\
NGC\,1340 & 032819.7$-$325555 &   3 &  2 & 1074 &  4 &   6 & 432 & 10.73 & 11.76 & 20.2 & 0.4 & 14.7 &  5 &  67 &  3 & 131\\
NGC\,1399 & 033230.3$-$360845 & 111 & 79 & 1411 & 23 & 244 & 454 & 12.30 & 13.94 & 19.6 & 0.2 & 19.3 &  0 &   0 &  7 & 116\\
IC\,1970  & 033631.5$-$440235 &   2 &  1 & 1090 &  9 &   9 & 103 & 10.06 & 10.06 & 19.4 & 2.0 & 14.9 &  8 &  90 & 15 &  99\\
NGC\,1386 & 033734.3$-$360610 &   8 &  5 &  755 & 26 &  70 & 165 & 10.37 & 12.40 & 19.8 & 0.9 & 10.3 &  1 &  58 &  7 & 113\\
NGC\,1398 & 033754.9$-$260905 &  10 &  4 & 1386 & 30 &  89 & 612 & 11.46 & 13.09 & 22.7 & 1.8 & 19.0 & 10 &  45 &  3 &  34\\
NGC\,1407 & 034002.6$-$191705 &  25 &  5 & 1713 & 34 & 167 & 385 & 11.61 & 13.22 & 24.2 & 1.2 & 23.5 & 17 &  47 & 10 &  47\\
NGC\,1395 & 034014.0$-$231718 &  24 & 11 & 1548 & 25 & 121 & 378 & 11.53 & 13.05 & 24.3 & 1.0 & 21.2 & 13 &  41 &  6 &  34\\
NGC\,1482 & 035438.9$-$212951 &   3 &  2 & 1715 & 55 &  78 &  48 & 10.70 & 11.32 & 28.9 & 3.1 & 23.5 & 16 &  28 &  9 &  23\\
NGC\,1519 & 040703.1$-$183633 &   4 &  2 & 1781 & 15 &  26 & 238 & 10.29 & 11.57 & 23.8 & 2.4 & 24.4 & 20 &  51 & 14 &  56\\
NGC\,1532 & 041226.4$-$332640 &  10 &  3 & 1159 & 30 &  89 & 137 & 11.25 & 12.70 & 20.3 & 1.5 & 15.9 &  9 &  72 &  9 &  98\\
\hhline{|-----------------|}
\end{tabular}}
\end{table*}


In the present paper we regard the dynamical center of the Fornax cluster
(NGC\,1399 group) situated near the core of the hot X-ray emitting gas (Jones et
al. 1997, Scharf et al. 2005, Machacek et al. 2005) as the gravity center of the
whole Fornax-Eridanus complex. Its spatial position was calculated as the mean
position of all NGC\,1399 group members, yielding $D = 19.6$~Mpc,
$\alpha=3^h32^m30^s$ and $\delta=-35^\circ51'15"$. For the moment this choice
(as the first approximation) seems to be reasonable since the NGC\,1399 region
has been investigated most detaily. The next step is to determine the spatial
position of the Fornax-Eridanus complex barycenter as the weighted mean for
position vectors of all the virialized substructures forming the complex. This
will be possible after obtaining more observational data in a wider area of the
complex. Still, the same calculation techniques are applied by us for both
center positions (with values $R=21.1$~Mpc, $\alpha=3^h33^m58^s$,
$\delta=-28^\circ44'45"$ adopted for the second case), and the resulting $R_0$
values didn't differ significantly. The barycenter position associated with
NGC\,1399 group yields 0.23~Mpc lower $R_0$ value from precise data and 0.19~Mpc
higher value from Tully-Fisher data, leading to the upper bound of discrepancy
$\sim 0.2$~Mpc. Generally speaking, the zero-velocity surface method is rather
stable in the sense of a barycenter position (Karachentsev \& Kashibadze 2006).

Table~2 represents groups, triplets and pairs forming the Fornax-Eridanus
complex with individual distance estimates (Karachentsev \& Makarov 2008;
Makarov \& Karachentsev 2009, 2011). Columns of the table contain: (1) name of
the brightest galaxy in a group/triplet/pair; (2) equatorial coordinates of the
brightest galaxy for triplets and pairs or mean equatorial coordinates for
groups (at the J2000.0 equinox); (3) number of galaxies in a system; (4) number
of galaxies with measured distances; (5, 6) mean velocity with respect to the
Local Group centroid and its error; (7) radial velocity dispersion; (8)
harmonical radius; (9) integrated luminosity of the group in the $K_s$ band;
(10) virial mass of the group; (11, 12) mean distance of the group and its
error; (13) {\itshape{}Hubble distance} calculated as $D_h = H_0 V_\mathrm{LG}$
where $H_0 = 73$~km~s$^{-1}$~Mpc$^{-1}$; (14) and (16) angular distance to the
Fornax-Eridanus complex barycenter for the cases of (14) dynamical center of the
NGC\,1399 group situated near the center of the hot X-ray emitting gas and (16)
dynamical center of all substructures forming the complex; (15) and (17) angle
between the directions to the Local Group and to the Fornax-Eridanus complex
barycenter for both cases mentioned above, respectively.

The possible role of barycenter position is illustrated by Figure~6 representing
velocity vs. distance diagrams where only the centers of the groups forming the
Fornax-Eridanus complex are shown. The top panel represents radial velocities
and distances of the groups relative to the Local Group centroid. The distance
to the Fornax-Eridanus complex, 21.1~Mpc, corresponds to the spatial position of
the complex barycenter as the weighted mean for position vectors of all the
virialized substructures forming the complex. The sets of dashed and dotted
lines indicate the Hubble flow perturbed by a point-like mass of
$2.16\cdot10^{14}M_{\odot}$ and $0.87\cdot10^{14}M_{\odot}$ respectively which
corresponds to the mass of the whole Fornax-Eridanus complex estimated in this
paper and the virial mass estimation of the Fornax cluster itself (Makarov \&
Karachentsev 2011). Both sets include S-shaped lines signing different angular
separations from the complex center ($0^\circ$, $5^\circ$, $15^\circ$ and
$30^\circ$). The error bars indicate rms uncertainties in distances and
velocities.

The velocity vs. distance diagram relative to the center of the Fornax-Eridanus
complex is shown in the bottom panel of the Figure~6. Two types of markers,
circles and squares, correspond to the center of the hot X-ray emitting gas and
the dynamical center of all substructures forming the complex, respectively. The
solid line indicates the Hubble flow perturbed by a point-like mass of
$0.87\cdot10^{14}M_{\odot}$ (the virial mass estimation of the Fornax cluster
itself) while the dashed and dotted lines are constructed as least square
estimates for both adopted center positions mentioned above
($0.53\cdot10^{14}M_{\odot}$ and $1.66\cdot10^{14}M_{\odot}$ respectively).

\begin{figure*}
\centering
\includegraphics[height=13cm,keepaspectratio,angle=270]{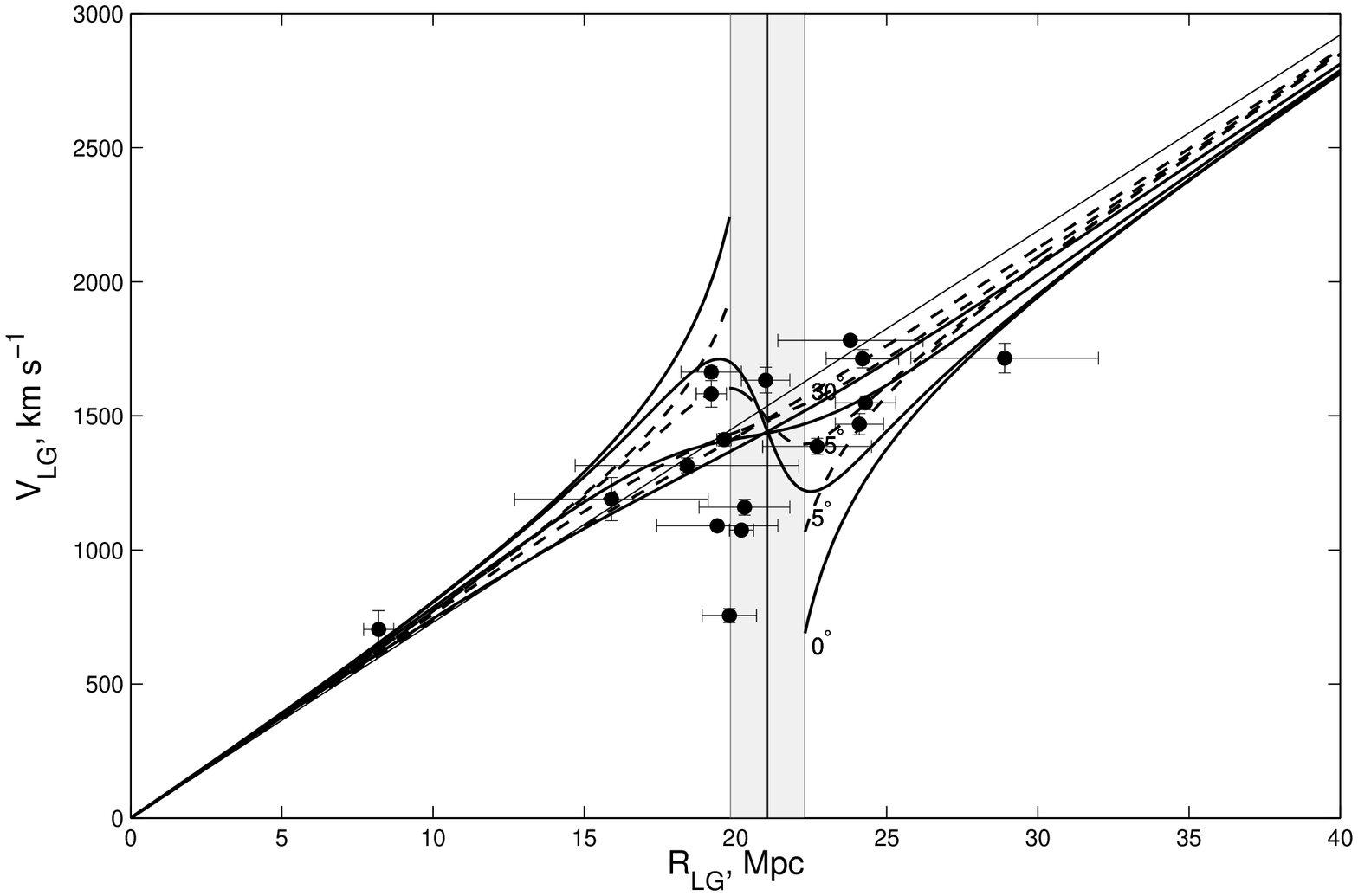}
\includegraphics[height=13cm,keepaspectratio,angle=270]{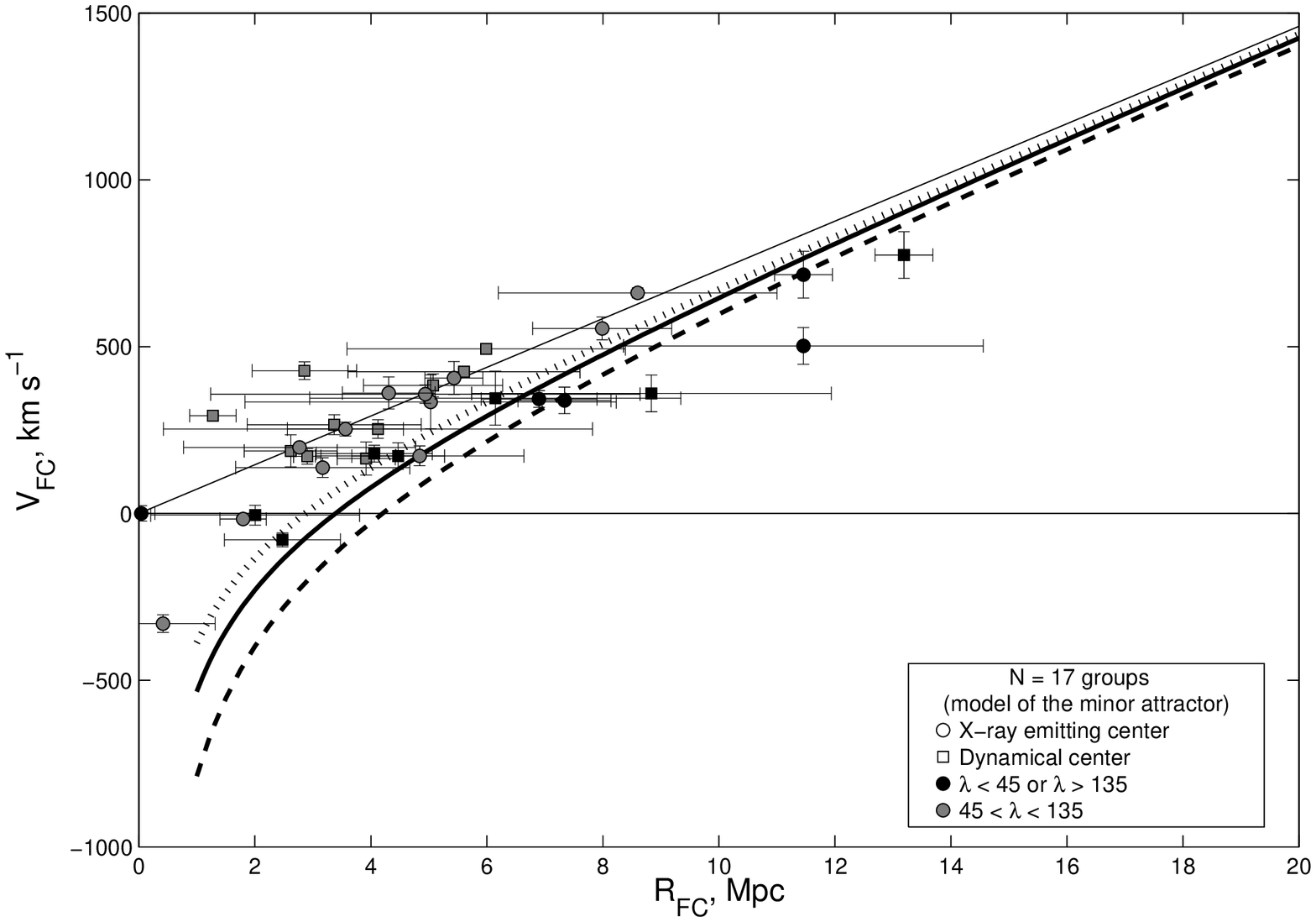}
\caption{{\itshape{}Top}: Radial velocities and distances of the groups relative
to the Local Group centroid. The distance to the Fornax-Eridanus complex,
21.1~Mpc, corresponds to the spatial position of the dynamical center of all
substructures forming the complex. The sets of dashed and dotted lines indicate
the Hubble flow perturbed by a point-like mass of $2.16\cdot10^{14}M_{\odot}$
and $0.87\cdot10^{14}M_{\odot}$ respectively, signing different angular
separations from the complex center ($0^\circ$, $5^\circ$, $15^\circ$ and
$30^\circ$). The error bars indicate rms uncertainties in distances and
velocities. {\itshape{}Bottom}: The velocity vs. distance diagram relative to
the center of the Fornax-Eridanus complex. Round and square markers correspond
to the center of the hot X-ray emitting gas and the dynamical center of all
substructures forming the complex, respectively. The dotted, solid and dashed
lines correspond to the Hubble flow perturbed by a point-like mass of
$0.53\cdot10^{14}M_{\odot}$, $0.87\cdot10^{14}M_{\odot}$, and
$1.66\cdot10^{14}M_{\odot}$ respectively.}
\end{figure*}

The structural complexity of the galaxy distribution suggests that probably the
best mass indicator for this cluster should be based on the velocity field of
the outskirt galaxies. Here two approaches will be adopted. In the first, the
radius of the zero velocity surface $R_0$ will be estimated from a running
median procedure. From the knowledge of $R_0$, the mass inside such a surface
follows immediately. In the second, the expected radial velocity profile derived
from the spherical model is fitted to the data and again, the mass inside the
zero-velocity surface results from the derived fit parameters.

\subsection{The running median}

The zero-velocity radius $R_0$ can be estimated from a running median procedure,
using directly observational data (Figure~7).

\begin{figure*}
\centering
\includegraphics[height=13cm,keepaspectratio,angle=270]{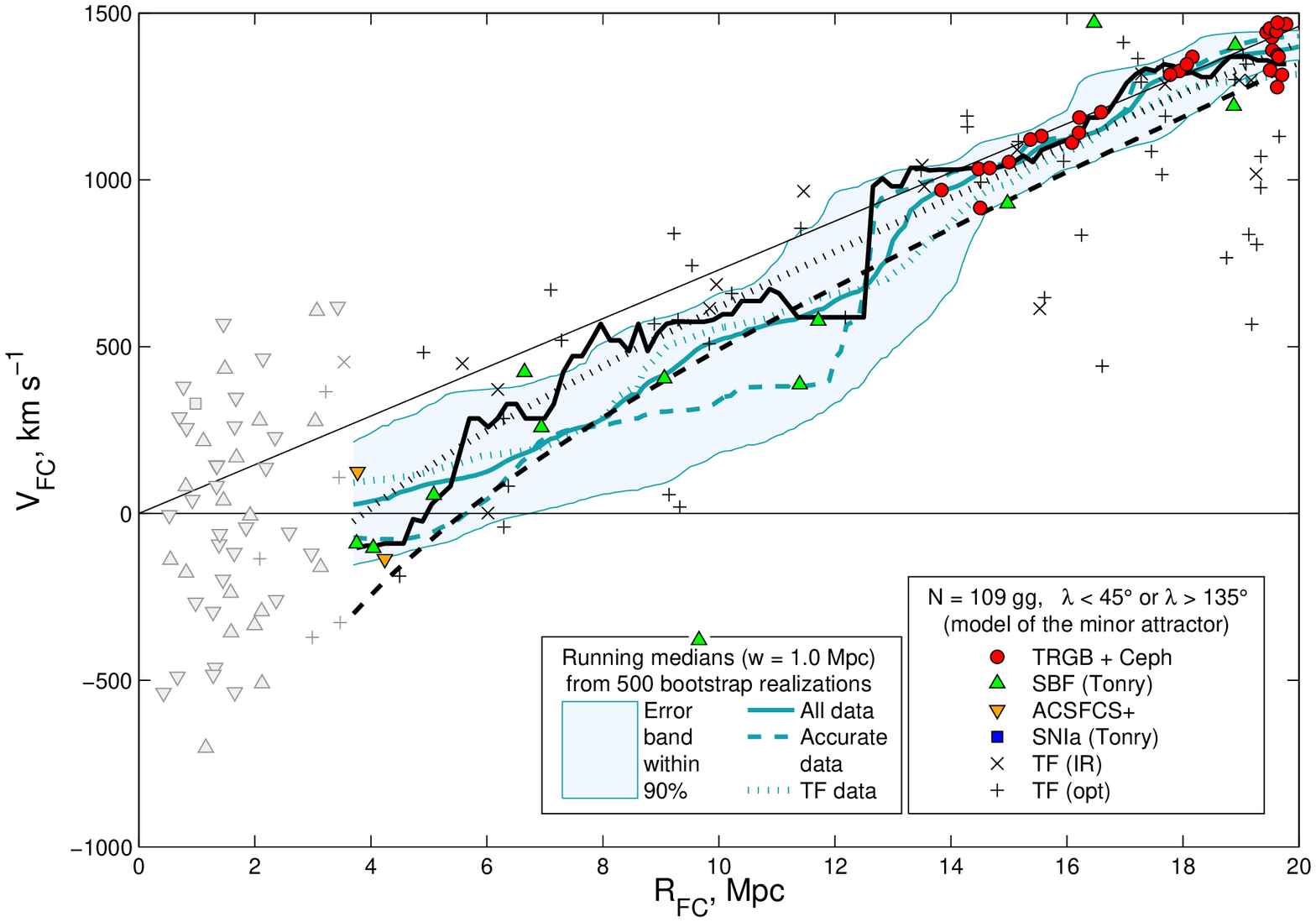}
\includegraphics[height=13cm,keepaspectratio,angle=270]{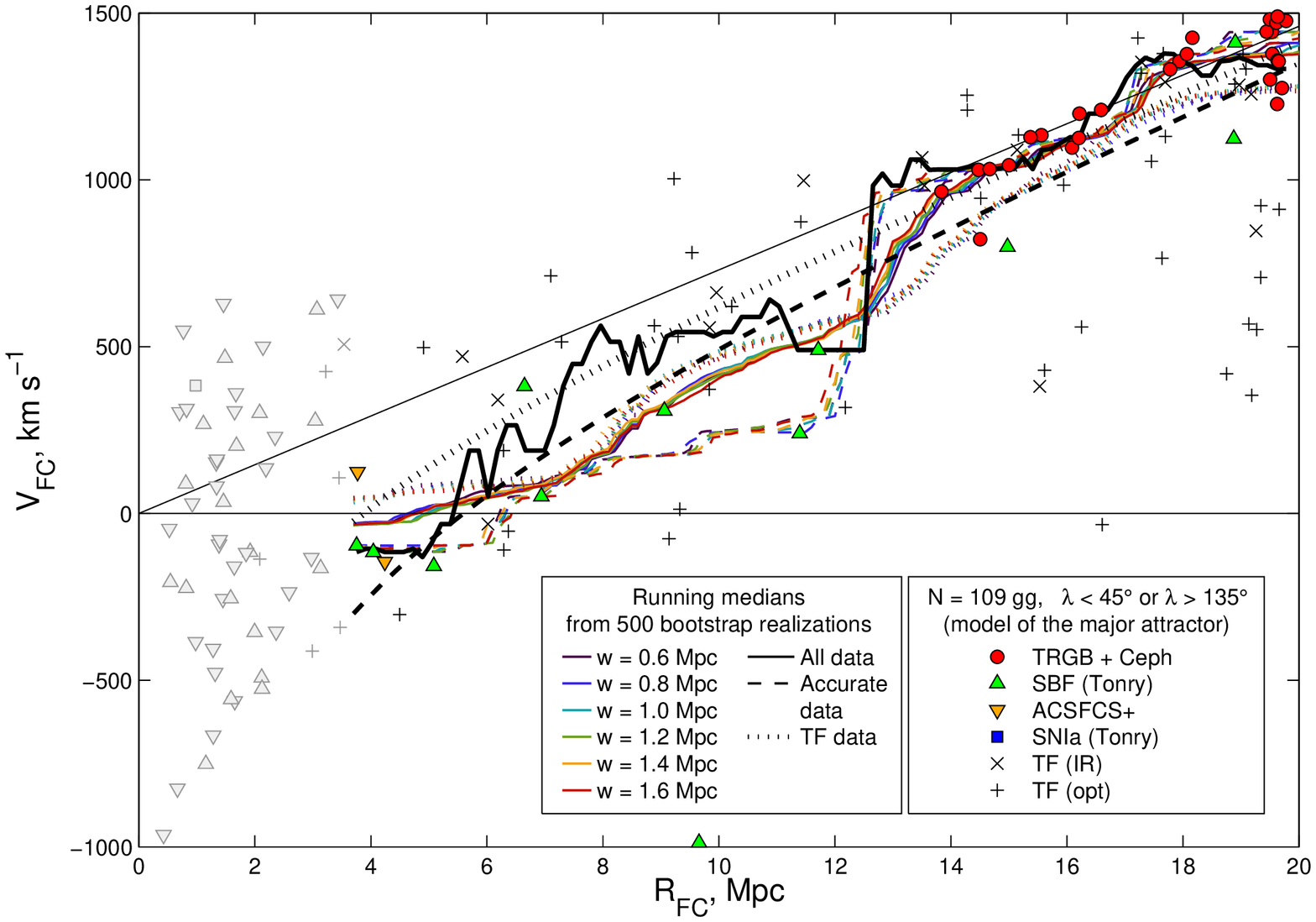}
\caption{Hubble flow in the Fornax cluster reference frame for 109 galaxies with
$R_\mathrm{FC}<20$ Mpc. The black solid polygon curve traces the running median
on observational data with a window of 1 Mpc. The dotted and dashed curves
correspond to a Hubble flow perturbed by masses of $1.30 \cdot 10^{14}M_{\odot}$
and $3.93 \cdot 10^{14}M_{\odot}$ respectively as the limiting cases within the
confidence range. Only the galaxies with $\lambda<45^\circ$ or
$\lambda>135^\circ$ are presented. {\itshape Top}: the case of almost pure
Hubble flow (``minor attractor'' approach). Blue curves trace the running
medians on simulated data with a window of 1 Mpc and the corresponding 90\,\%
error band. {\itshape Bottom}: the case of almost pure infall towards the Fornax
cluster (``major attractor'' approach). Coloured curves trace the running
medians on simulated data with different window values.}
\end{figure*}

In order to account errors in observed distances and velocities and to give some
estimates of the uncertainties, a Monte-Carlo simulations technique was used.
For any galaxy $i$ in a given dataset $j$, a corresponding distance from the
center is generated according to the relation $R_i = R_{i,obs} + \Delta R_i
\beta_{ij}$, where $R_{i,obs}$ is the observed distance from the Fornax center,
$\Delta R_i$ is an observational error associated to the distance and
$\beta_{ij}$ is a normally distributed random number (with $\sigma=1$).
Generated velocities were derived by a similar procedure. About 10,000 datasets
were generated and for each of them the running median method was applied,
yielding different values of $R_0$, which were then averaged. The distribution
of 10,000 $R_0$ are shown in Figure~8 as well as the mean and median values and
the 90\,\%error band (for the window $w=1$~Mpc). The resulting median values and
corresponding errors for (A) minor attractor and (B) major extended attractor
cases are given in Table~3. The first two columns correspond to the cases when
only precise or Tully-Fisher data were used for Monte-Carlo simulations while
the third column corresponds to the whole dataset. The lines of the table
correspond to the different median windows.

\begin{figure*}
\sidecaption
\includegraphics[height=12cm,keepaspectratio,angle=270]{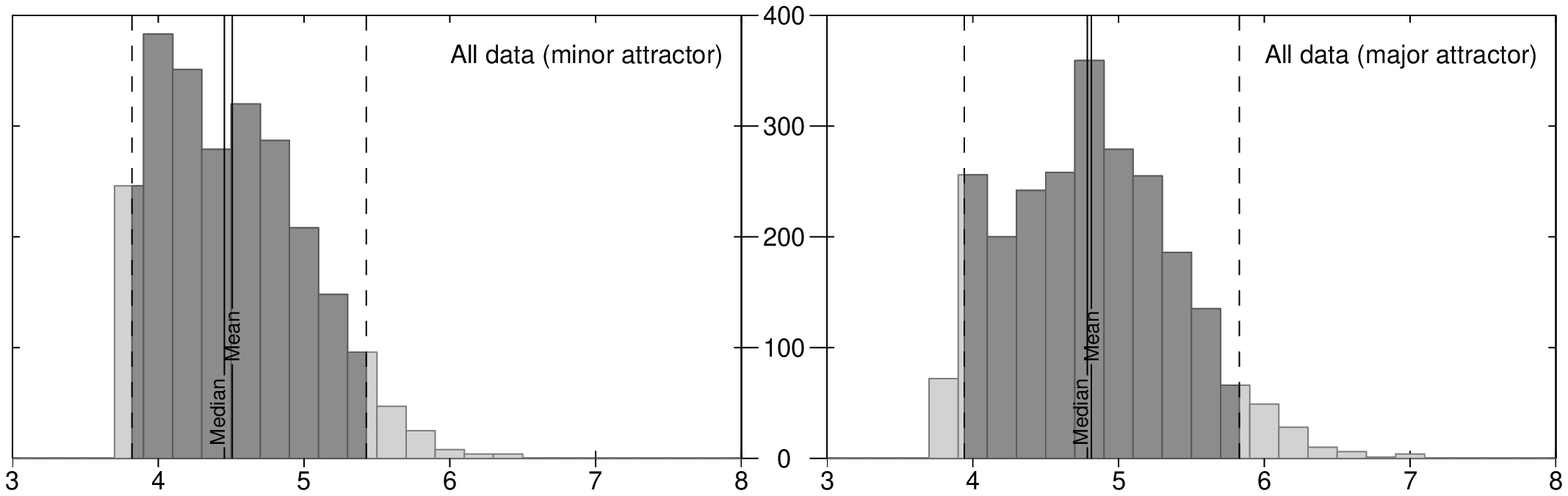}
\caption{Distribution of 10,000 $R_0$ realizations simulated using Monte-Carlo
technique, their mean and median values and the 90\,\% error band. The window
used for the running median procedure is $w=1$~Mpc.
\vspace{-3.7cm}}
\end{figure*}

\begin{table*}
\centering
\caption{Radius $R_0$ and its 90\,\% error band (in Mpc) obtained from
Monte-Carlo simulations.}
\begin{tabular}{|c|cc|cc|cc|}
\hhline{|-|--|--|--|}
\multicolumn{1}{|c|}{Window}&    
\multicolumn{2}{c|}{Precise distances}&
\multicolumn{2}{c|}{TF distances}&
\multicolumn{2}{c|}{All}\\
\multicolumn{1}{|c|}{(Mpc)}&
\multicolumn{2}{c|}{(samples 1--4) }&
\multicolumn{2}{c|}{(samples 5--6) }&
\multicolumn{2}{c|}{samples}\\
\hhline{|~|--|--|--|}
\multicolumn{1}{|c|}{ }&
\multicolumn{1}{|c}{(A)}&
\multicolumn{1}{c|}{(B)}&
\multicolumn{1}{c}{(A)}&
\multicolumn{1}{c|}{(B)}&
\multicolumn{1}{c}{(A)}&
\multicolumn{1}{c|}{(B)}\\
\hhline{|-|--|--|--|}
0.8&     4.48    &     4.56    &     4.87     &     5.15     &     4.41    &     4.67    \\
   & 3.88$-$5.46 & 3.94$-$5.63 & 3.95$-$6.18  & 4.08$-$6.68  & 3.83$-$5.39 & 3.93$-$5.76 \\
\hhline{|-|--|--|--|}
1.0&     4.58    &     4.72    &     5.04     &     5.39     &     4.42    &     4.78    \\
   & 3.87$-$5.50 & 3.95$-$5.75 & 3.95$-$6.32  & 4.09$-$6.81  & 3.82$-$5.42 & 3.93$-$5.79 \\
\hhline{|-|--|--|--|}
1.2&     4.51    &     4.66    &     5.15     &     5.51     &     4.39    &     4.83    \\
   & 3.88$-$5.56 & 3.92$-$5.85 & 3.95$-$6.33  & 4.16$-$6.86  & 3.82$-$5.48 & 3.93$-$5.88 \\
\hhline{|-|--|--|--|}
\end{tabular}
\end{table*}


With the Fornax cluster distance of $\sim20$~Mpc and the typical uncertainty of
10\,\% for the ACSFCS+ galaxies populating the central core of the cluster, the
observational distance errors in the virialized zone will be of about 2~Mpc.
Assuming the virial radius of the Fornax cluster to be $\sim1.5$ Mpc we should
conclude that the cluster core galaxies can possibly distort the pattern in the
$R_0$ region for the Fornax cluster itself, but not for the whole
Fornax-Eridanus complex. Anyway, the galaxies in the Fornax cluster core, i.e.
those with $R_\mathrm{FC}<3.5-3.7$~Mpc, were not regarded in the Monte-Carlo
data simulations, resulting to 109 galaxies (see Figure~7).

The substantial dip in the running median of simulations based on precise data
can be explained partially by the enormous velocity of NGC\,1400 (see, for
example, Perrett et al. 1997). Belonging to the NGC\,1407 group, NGC\,1400 has
an extremely low velocity relative to the Local Group (558~km~s$^{-1}$) while
NGC\,1407 has $V_\mathrm{LG} = 1771$~km~s$^{-1}$. Both galaxies have roughly the
same distance (26.4~Mpc for NGC\,1400 and 28.8~Mpc for NGC\,1407 according to
Tonry et al. 2001). In the velocity-distance diagrams NGC\,1400 has a
significant negative velocity relatively to the Fornax-Eridanus complex center,
$-$380~km~s$^{-1}$ and $-$988~km~s$^{-1}$ for the cases of minor and major
attractors respectively, and therefore it appears in the legend region at the
Fornax-centric distance of 9.6~Mpc shifting the running median downwards.
 
According to the Table~3, the median estimate of $R_0$ is 4.60~Mpc with
confidence interval corresponding to 90\,\% error band of
$[3.88-5.60]$~Mpc.

Once $R_0$ is known, the mass inside the zero-velocity surface can be computed.
Using the spherical model, including the effects of the cosmological constant,
the mass inside $R_0$ is (E. Shaya, private communication; Karachentsev et al.
2007)
\begin{equation}
M_T = \frac{\pi^2}{8G} R_0^3 \frac{H^2_0}{f^2(\Omega_m)}
\end{equation}
\noindent{}where
\begin{equation}
f(\Omega_m) = \frac{1}{1-\Omega_m} - \frac{\Omega_m}{2(1-\Omega_m)^{3/2}}\cdot \mathrm{arccosh}(\frac{2}{\Omega_m}-1).
\end{equation}
with $\Omega_m$ being the mass density parameter.

Figure~9 shows the ratio of masses of a galaxy system computed in the
$\Lambda$CDM model and in the empty Universe model respectively as a function of
$\Omega_m$. As one can see, the adopted uncertainty in $\Lambda$ value affects
as about 3\,\% in mass estimation, which is negligible as compared with
uncertainties caused by observational errors.

\begin{figure}
\resizebox{\hsize}{!}{\includegraphics[angle=270]{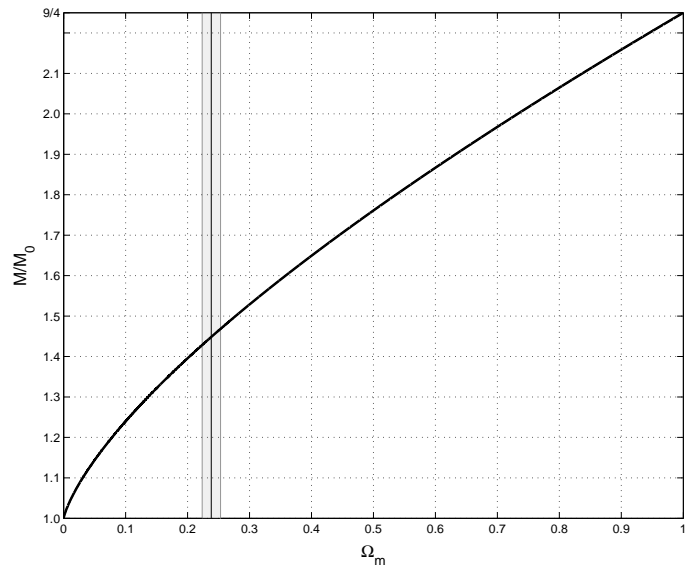}}
\caption{Ratio of masses of a galaxy system computed in the $\Lambda$CDM model
and in the empty Universe model respectively as a function of $\Omega_m$. The
value of $\Omega_m = 0.238 \pm 0.015$ adopted in the $\Lambda$CDM model is
marked with a vertical gray strip.}
\end{figure}

$\Lambda$CDM parameters can be determined from WMAP data with a sufficient
accuracy, i.e., $H_0 = 73.2^{+3.1}_{-3.2}$~km~s$^{-1}$~Mpc$^{-1}$ and $\Omega_m =
0.238\pm0.015$ (Spergel et al. 2007). Substituting these values into the
Equation~(1), we get
\begin{equation}
(M_T/M_{\odot})_{0.24}=2.23\cdot10^{12}(R_0/\mathrm{Mpc})^3.
\end{equation}
Notice that from the numerical solution of the equations describing the
Lema\^itre-Tolman model, modified in order to include effects of the
cosmological constant, Peirani \& de Freitas Pacheco (2008) obtained for the
mass inside the zero-velocity surface
\begin{equation}
M = 4.24\times 10^{12}h^2R_0^3 ~~M_{\odot}
\label{eq4}
\end{equation}
The numerical coefficient was obtained for $\Omega_m = 0.3$ and for $h = 0.73$
both relations derived either analytically or numerically agree quite well. For
$R_0$ = 4.60~Mpc the corresponding total mass of the Fornax complex is $M = 2.16
\times 10^{14} M_\odot$ and for the limiting values of 90\,\% error band
$[3.88-5.60]$~Mpc we obtain $M = [1.30-3.93] \times 10^{14} M_\odot$ as a
confidence interval.

\subsection{Mass estimate from the radial velocity profile}

In this section, the mass inside the zero-velocity surface is estimated by
fitting a theoretical profile directly to data. We follow the procedure
developed by Peirani \& de Freitas Pacheco (2006, 2008), who have numerically
computed the velocity field of outskirt galaxies, based on the spherical
collapse model including effects of the cosmological constant. This approach
assumes that: a) most of the cluster mass is concentrated in the core, in which
the shell crossing has already occurred and b) that orbits of galaxies outside
the core are mainly radial. From the numerical calculations by Peirani \& de
Freitas Pacheco (2008), the velocity-distance relation is
\begin{equation}
\label{modelvelocity}
V(R) = 1.377H_0R - \frac{0.976H_0}{R^n}\left(\frac{GM}{H^2_0}\right)^{(n+1)/3}.
\end{equation}

Here $M$ is the core cluster mass, $R$ is the distance of the member galaxy to
the cluster center, $V(R)$ is the radial velocity of the galaxy with respect to
the mass center, $H_0$ is the present value of the Hubble parameter and
$n=0.627$. The relation above results from a fit of numerical data and is valid
for the present time ($z = 0$), since it varies as a function of the redshift.
Notice that eq.~\ref{eq4} can be deduced from this equation when $V(R_0) = 0$.

\begin{figure*}
\centering
\includegraphics[height=13cm,keepaspectratio,angle=270]{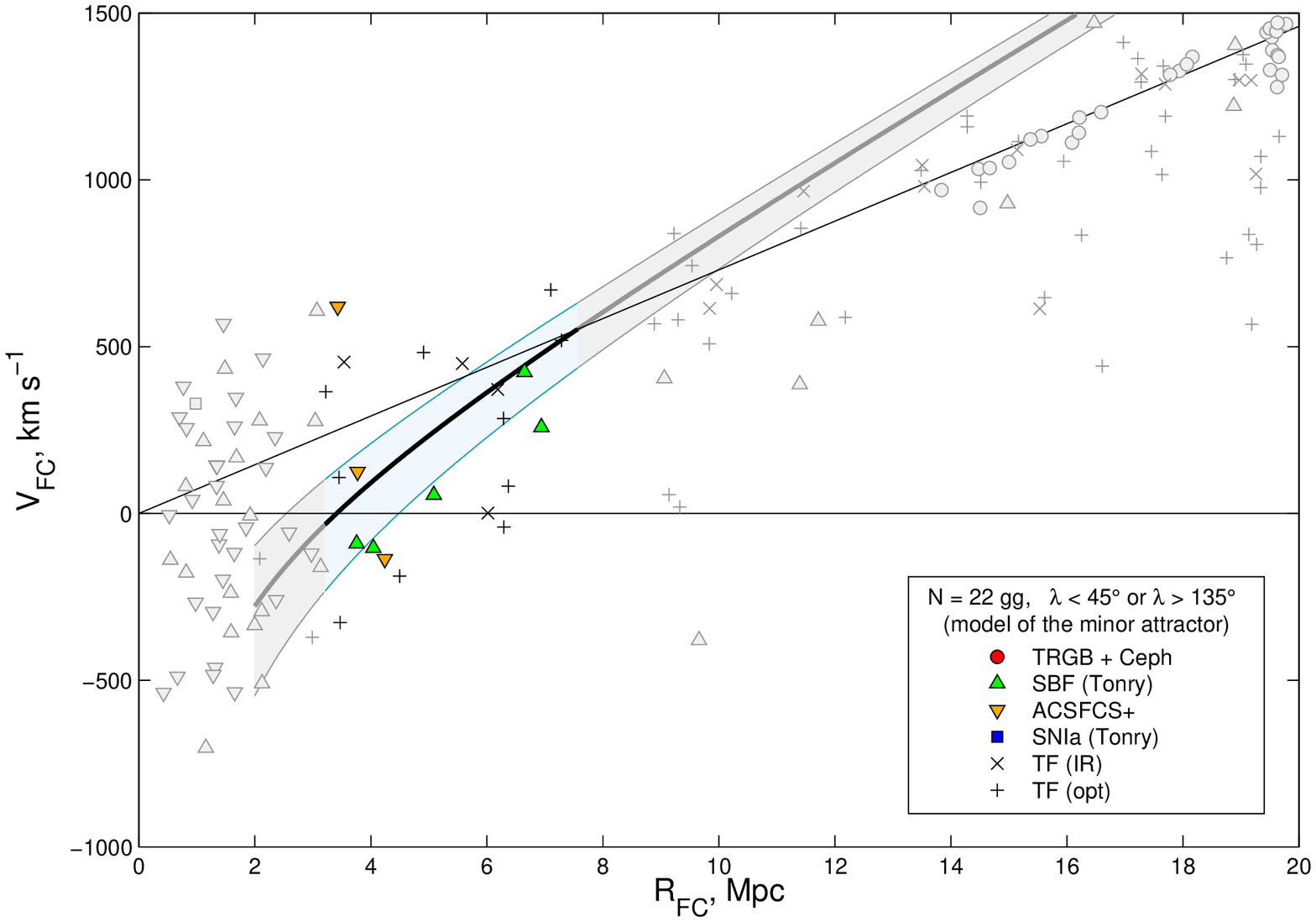}
\includegraphics[height=13cm,keepaspectratio,angle=270]{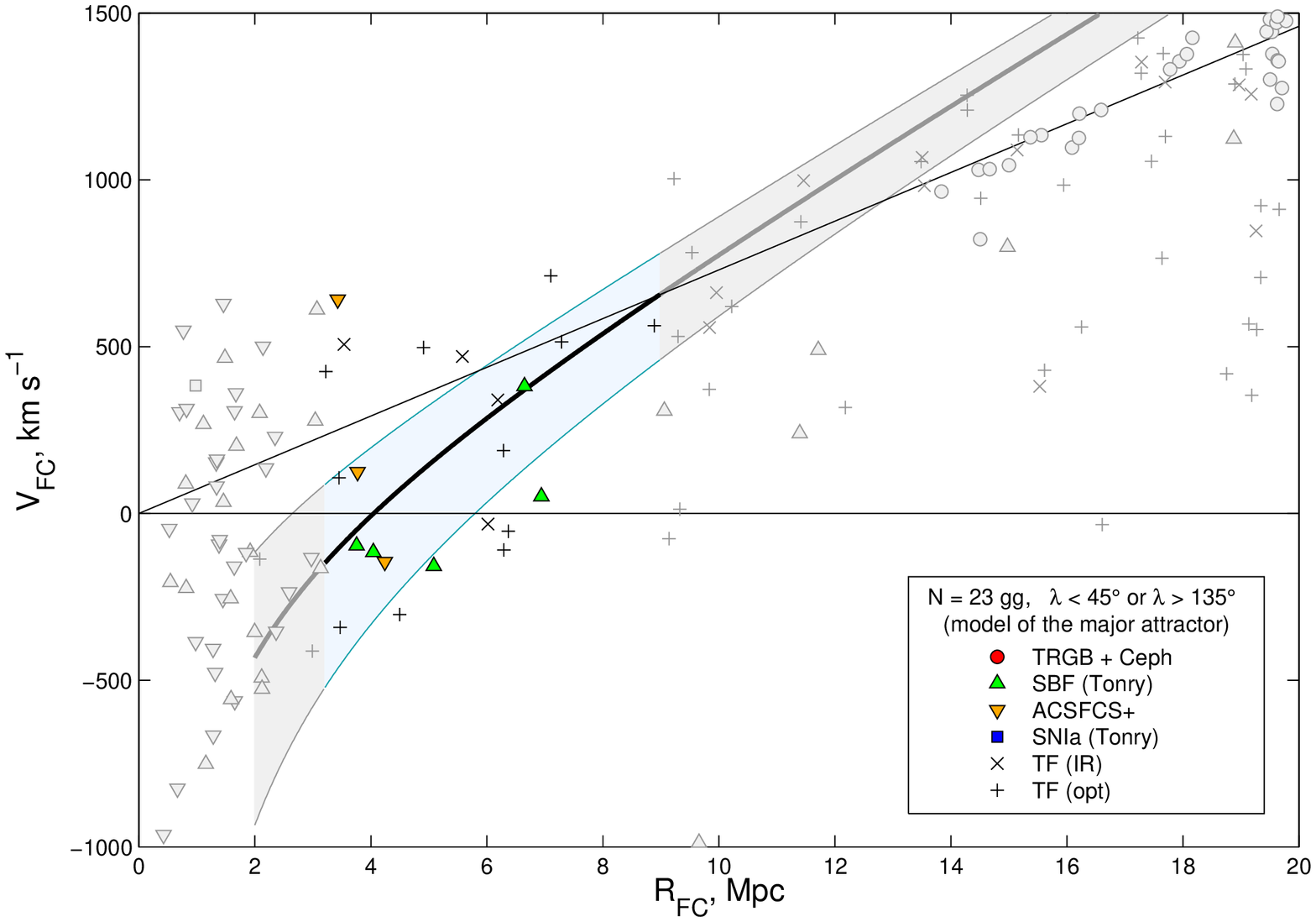}
\label{fornaxreffr}
\caption{Velocity field in the Fornax region, for 164 galaxies with distances
less than 20 Mpc from the cluster centre. The 90\,\% error band resulting from
Monte-Carlo simulations is shown. {\itshape Upper panel}: velocities derived
according to the ``minor attractor'' approach. {\itshape Lower panel}:
velocities derived according to the ``major attractor'' model.}
\end{figure*}

In principle, if accurate radial velocities and distances are available, from
the fitting of eq.~\ref{modelvelocity} to data it is possible to derive both the
core cluster mass and the Hubble parameter (Peirani \& de Freitas Pacheco 2008).
Here face to the uncertainties in the determination of Fornax-centric
velocities, as explained in the previous section, we adopted the procedure of
fixing $H_0$ as 73~km~s$^{-1}$~Mpc$^{-1}$ and computing the mass either for the
``minor attractor'' or for the ``major extended attractor'' models.

\begin{table}[]
\caption{Masses resulting from the fit of eq.~\ref{modelvelocity} to
the present data.}
\begin{center}
\begin{tabular}{lcc}
\hline
Parameter &``Minor attractor'' &``Major attractor'' \\
\hline
Mass (in $10^{14}~M_{\odot})$&    0.93   &    1.55   \\
                             &0.38$-$2.10&0.43$-$4.54\\
$R_0$ (in Mpc)               &    3.44   &    4.08   \\
                             &2.57$-$4.52&2.67$-$5.85\\
$\sigma$ (in km~s$^{-1}$)    &    300    &    345    \\
$\sigma_c$ (in km~s$^{-1}$)  &     84    &    190    \\
\hline
\end{tabular}
\end{center}
\end{table}

Masses resulting from our fitting procedure are given in Table~4 for the two
models adopted to estimate Fornax-centric velocities. The zero velocity surface
radius $R_0$ and the velocity dispersion $\sigma$ with respect to the mean
infall flow are also given. The second column corresponds to velocities
estimated from the ``minor attractor'' approach whereas the third column gives
values derived from the ``major extended attractor'' model. The corresponding
confidence intervals for mass and $R_0$ values are estimated as 90\,\% error
band. The third and the fourth lines give respectively the uncorrected and the
corrected velocity dispersions with respect to the infall flow towards Fornax.
Notice that this method leads to values of masses that are, on the average, 50\%
smaller than the ``running median'' approach, being rather a mass estimate for
the Fornax cluster core than for the total Fornax-Eridanus complex. It is worth
mentioning that the derived 1-D velocity dispersion with respect the bulk
motion, (300--345~km~s$^{-1}$), is one order of magnitude higher than that
observed in the vicinities of the Local Group but it drops to
(84--190~km~s$^{-1}$) when the distance measurement errors are taken into
account. Figure~10 shows the best fit between the expected velocity profile
(eq.~\ref{modelvelocity}) derived from the spherical model and data, whose
velocities were computed according to the ``minor attractor'' model (upper
panel) or to the ``major attractor'' model (lower panel). The number of galaxies
used in the fitting procedure is limited by $R_\mathrm{max}$ satisfying the
relation $R_\mathrm{max}H_0 = V(R_\mathrm{max})$ and actually accounts for 22
and 23 galaxies respectively.

Considering the mass values estimated from these procedures we adopt for the
Fornax cluster itself a mass of $1.24 \times 10^{14}~M_{\odot}$ with the
confidence interval of $[0.40-3.32]\times 10^{14}~M_{\odot}$, and for the total
Fornax-Eridanus complex a mass of $2.16\times 10^{14}~M_{\odot}$ with the
confidence interval of $[1.30-3.93]\times 10^{14} M_\odot$, which corresponds to
a value half order of magnitude the Virgo cluster mass. In Table~5 we compare
previous mass estimates for the Fornax cluster (normalized to $D_{For} = 20$~Mpc
and $D_{Eri} = 25$~Mpc) with the present study, based on the velocity field of
outskirt galaxies.

\begin{table*}
\centering
\caption{Virial and total mass estimates (in $10^{14} M_\odot$ units) for the
Fornax-Eridanus complex.}
\begin{tabular}{|c|c|c|c|l|}
\hhline{|-|-|-|-|-|}
Fornax  & Eridanus & Other clumps &  Total & Reference                         \\
\hhline{|-|-|-|-|-|}
1.53       &   0.20   &  0.15  &    1.88   &1                     \\
1.15       &   0.70   &  $-$   &    1.85   &2                     \\
$-$        &   0.7    &  $-$   &    $-$    &3                     \\
1.00       &   0.92   &  $-$   &    1.92   &4                     \\
0.87       &   0.62   &  0.43  &    1.92   &5                     \\
\hhline{|-|-|-|-|-|}
0.40$-$3.32&    $-$   &  $-$   &    $-$    &6, eq.~\ref{modelvelocity}\\
 $-$       &    $-$   &  $-$   &1.30$-$3.93&6, via $R_0$              \\
\hhline{|-|-|-|-|-|}
\end{tabular}
\tablebib{
(1) Tully 1987; (2) Ferguson \& Sandage 1990; (3) Brough et al. 2006;
(4) Crook et al. 2007; (5) Makarov \& Karachentsev 2011; (6) this paper.}

\end{table*}

\section{Concluding remarks}

The distribution of galaxies in the vicinity of the Fornax cluster indicates a
substantial degree of subclustering that forming the Fornax complex. As a
consequence, mass estimates based on the virial relation are affected by the
fact that the system has not yet reached an equilibrium state and is still
probably in formation (Dunn \& Jerjen 2006). In fact dwarf galaxies have
distinct dynamic properties in comparison with bright galaxies and are probably
infalling into the system (Drinkwater et al. 2001). These difficulties may be
circumvent by studying the velocity field of outskirt galaxies, which permits an
estimate of the zero-velocity surface and, consequently of the mass inside such
a surface as proposed originally by Lynden-Bell (1981) and Sandage (1986).

In this investigation, a culled sample of 109 galaxies with measured distances
and within 20 Mpc from the cluster center were used to study the velocity field
in the neighborhood of the Fornax cluster. Since tangential velocities are
unknown, in order to estimated the galaxy velocities with respect to the cluster
center two main assumptions were made: in the first, it was supposed that the
velocity vector is essentially dominated by the Hubble flow (``minor attractor"
model) while in the second, it was assumed that galaxies are infalling radially
(``major attractor" model). The zero-velocity radius was derived by two
different methods: the ``running median" and by fitting directly the data to the
expected velocity profile derived from the spherical model, including the
effects of a cosmological constant.

The best fit radius of the zero-velocity sphere for the Fornax-Eridanus complex
is estimated by us to be $R_0 = 4.60$~Mpc with the confidence interval of
$[3.38-5.60]$~Mpc while the mass inside such a surface is $M_{tot} = [1.30-3.93]
\times 10^{14}~M_{\odot}$. At the distance to Fornax cluster 20~Mpc, the radius
$R_0 = 4.60$~Mpc corresponds to $13.2^\circ$ shown in Figure~5. Notice that
within this circle there are almost all systems identified by Makarov \&
Karachentsev (2011) as virialized groups bound to Fornax cluster. Their total
virial mass, $M = 1.92 \times 10^{14}~M_{\odot}$, agrees with $M_{tot}$, meaning
that probably only a small part of the Fornax-Eridanus complex mass is spreaded
between the groups.

\begin{acknowledgements}
Support of this work has been provided by Henri Poincar\'e Junior Fellowship of
ADION in 2010, Observatoire de la C\^ote d'Azur, CNRS, France; the Dynasty
Foundation of Noncommercial Programs, Russia; The Ministry of Education and
Science of the Russian Federation (contract no. 14.740.11.0901); the Russian
Foundation for Basic Research (projects no. 10-02-92650, 11-02-00639).
\end{acknowledgements}

{}

\end{document}